\documentclass[aps,pre,reprint,amsmath,amssymb,showpacs]{revtex4-1}
\usepackage{graphicx}
\usepackage{epstopdf}
\usepackage[utf8x]{inputenc}
\usepackage{color}

\begin{document}
\title[]{Dynamical analysis of an optical rocking ratchet: Theory and experiment}

\author{Alejandro \surname{V. Arzola}}
\email[]{avarzola@gmail.com}
\affiliation{Institute of Scientific Instruments of the ASCR, v.v.i., Academy of Sciences of the Czech Republic, Kr\'alovopolsk\'a 147,
612 64 Brno, Czech Republic}
\author{Karen \surname{Volke-Sep\'ulveda}}
\email[]{karen@fisica.unam.mx}
\affiliation{ICFO-Institut de Ci\`encies Fot\`oniques, Mediterranean Technology Park, 08860 Castelldefels, Barcelona, Spain. Permanent address: Instituto de F\'isica, Universidad Nacional Aut\'onoma de M\'exico, Apartado Postal 20-364, 01000 M\'exico, Distrito Federal, Mexico}
\author{Jos\'e \surname{L. Mateos}}
\email[]{mateos@fisica.unam.mx}
\affiliation{Instituto de F\'isica, Universidad Nacional Aut\'onoma de M\'exico, Apartado Postal 20-364, 01000 M\'exico, Distrito Federal, Mexico}

\begin{abstract}
A thorough analysis of the dynamics in a deterministic optical rocking ratchet (introduced in A. V. Arzola \textit{et al.}, Phys. Rev. Lett. \textbf{106,} 168104 (2011)) and a comparison with experimental results are presented. The studied system consists of a microscopic particle interacting with a periodic and asymmetric light pattern, which is driven away from equilibrium by means of an unbiased time-periodic external force. It is shown that the asymmetry of the effective optical potential depends on the relative size of the particle with respect to the spatial period, and this is analyzed as an effective mechanism for particle fractionation. The necessary conditions to obtain current reversals in the deterministic regime are discussed in detail.   
\end{abstract}
\maketitle

\section{Introduction}

The phenomenon of directed transport in non-linear dynamical systems is a very active research field, since it has relation with many processes in nature. This includes the so-called ratchet effect. The key ingredients for this effect to take place are: a periodic potential, a symmetry breaking, either in space or in time or both, and an unbiased time-modulated external driving mechanism (the non-equilibrium condition). The many different ways to incorporate these ingredients, as well as the different dynamic conditions (inertial or overdamped, deterministic or stochastic) lead to a huge variety of models grouped under the same general appellation, because they all produce the same general result: a rectified motion under the action of unbiased forces. This model has been successfully applied to describe the transport of molecular motors inside the eukaryotic cells, cold atoms in optical lattices, vortices in superconducting devices, granular flows, and cell migration, among many others \cite{hanggi_artificial_2009, mahmud_directing_2009,Renzoni2003,RGommers2006,Renzoni2008,Linke_Self-propelled_2006,Villegas_Science_2003,marquet_rectified_2002,matthias_asymmetric_2003,Visscher_1999,denisov_ac-driven_2009}. This intriguing mechanism has unique properties not fully understood, nor exploited, such as the current reversals \cite{mateos_chaotic_2000,mateos_2003}. Due to their great interest, ratchets keep being a very appealing subject from both the theoretical and the experimental viewpoints.

In particular, the optical micromanipulation techniques have become ideal tools to study the paradigmatic behavior of microscopic particles in periodic and quasiperiodic potentials \cite{ashkin_optical_2006,Bohlein_2012, gopinathan_statistically_2004, Xiao_2010, Bechingerkinks2012, Evstigneev_diffusion_2008, dholakia_shaping_2011,Mu_enhancedparticletransport_2009,ricardez_sieve_2006, MacDonald_Microfluidic_2003}. They offer a high degree of control and versatility to perform both model experiments and practical application devices. Many configurations of light distribution have been explored using discrete arrays of optical traps or continuous interfering patterns of light. In the majority of cases, the microscopic particles are driven by an external drag force through the light field, resulting in interesting transport mechanisms. Examples are the fractionation of particles according to their size \cite{gopinathan_statistically_2004}, or formation of collective entities of particles within the lattice, such as kinks and antikinks \cite{Bechingerkinks2012}, explained by the particle size-dependence of the optical force on the lattice period and/or from their collective behavior in the lattices. These optical micromanipulation experiments lie in the overdamped regime, and most of them have been done with symmetric lattices, but there are also interesting realizations of ratchet systems. Particularly, the first experimental implementation of a thermal ratchet with an optical trap was due to Faucheux and coworkers \cite{Faucheux_PhysRevLett}. In that case, an asymmetric light pattern was created with an individual trap rapidly scanning over a circumference and a time modulation was introduced via an on-off mechanism, also known as pulsating or flashing driving. The rectification of the Brownian motion of a single particle was observed. Later realizations of optical ratchets were also based on spatially periodic potentials created with holographic optical tweezers and a flashing driving mechanism as well \cite{xiao_two-dimensional_2011,lee_one-dimensional_2005,SHyuk}. In the case of a flashing driving, the transport relies on thermal diffusion, so it is possible only for Brownian particles. 

In contrast, more recently we reported the first demonstration of a non-Brownian microscopic ratchet using optical micromanipulation \cite{arzola2011experimental}. This was based on a rocking driving mechanism that pushes the particle in alternating directions through a stationary, asymmetric and periodic pattern of light. There are two novel aspects to address in this realization, namely, the particles are in the range of sizes where the Brownian motion is negligible (deterministic regime), but it is still in an overdamped regime, and the external rocking driving is discontinuous in time. Specifically, the time-periodic rocking force is a three-state function; it alternates between semi-cycles of constant positive and negative values (in which the force tilts the potential in opposite directions) with a waiting-time between them in which the force is zero. In this paper, we present a detailed study and new results considering the optical force and the dynamics of a deterministic optical rocking ratchet. We start by discussing the particle-dependence of the optical potential, a fact rarely considered in theoretical ratchet models, but very important in real systems. The possibility of using this effect to design a sensitive fractionation device is also discussed. Then we present a thorough analysis of the dynamics of the system within a general scope. We investigate the role of the finite waiting-time in the rocking force for the arising of the current reversals phenomenon, observed in our previous work \cite{arzola2011experimental}. Finally, we establish a quantitative comparison between the theoretical and experimental results in some specific cases.

\section{Optical Rocking Ratchet}\label{sec_rocking}

\subsection{Optical Ratchet Potential}

It is well known that the optical force resulting from the interaction of a light field with a microscopic dielectric particle has two contributions: the scattering force and the gradient force \cite{ashkin_optical_2006}. Whereas the former always pushes the particle along the direction of propagation of the light field, the latter is a conservative contribution whose magnitude is associated with the intensity gradient. 

\begin{figure}
\includegraphics[scale=0.5]{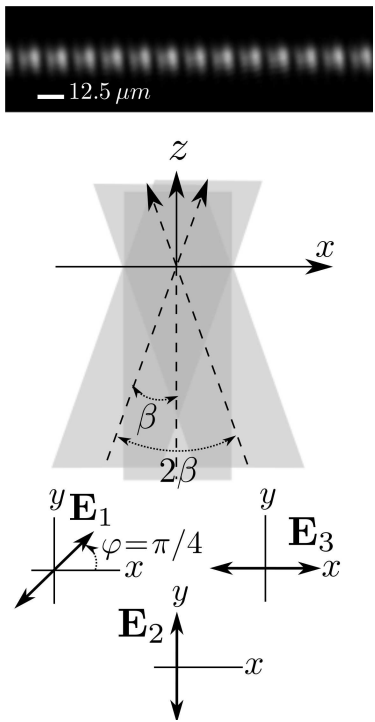}
\caption{Schematic of the interference by pairs of three linearly polarized beams with polarizations along the indicated directions. This gives rise to two superimposed patterns of fringes, one of them with twice the period of the other. In the top of the figure is shown an experimental picture that exemplify the resulting intensity pattern in the experiment.}\label{fig_3beams}
\end{figure}

A ratchet potential, which is periodic but asymmetric, was generated by overlapping two patterns of fringes with a sinusoidal profile, one of them with twice the period of the other and mutually orthogonal polarizations. This was achieved by interfering three light beams by pairs, controlling their relative polarization states, intensities and phases, as explained in Ref.~\cite{arzola2011experimental} (see Fig.~\ref{fig_3beams}). In the set of experiments discussed here, the light distribution propagates upwards along the vertical direction ($z$-axis). The pattern is widely extended along the direction of its periodicity ($x$-axis), while being very narrow in the orthogonal direction ($y$-axis), giving rise to a quasi-1D dynamics \cite{arzola2011experimental, Demergis:11}. The weight of the particles is large enough to overcome the scattering optical force along the $z$ axis. Therefore, we will focus our analysis on the gradient optical force along the $x$ axis.

It has been extensively demonstrated before that the gradient optical force exerted on a spherical dielectric particle of radius $R_0$ by a single sinusoidal pattern of fringes has the same periodicity, but its magnitude depends strongly on the relative size of the particle with respect to the period of the light lattice \cite{jonas2008light,arzola2009force,dholakia2007optical,dholakia2007cellular,ricardez_sieve_2006} . This behavior is the basis of many optical sorting devices \cite{jonas2008light}. In the case of the superposition of two periodic patterns of fringes with orthogonal polarization, the total gradient force acting on the sphere can be expressed as the sum of the forces exerted by each of the individual patterns, namely, 
\begin{align}\label{eq_fuerza}
F(x;\Lambda,R_0)=\frac{P}{c}&\left[A_{\perp}(\Lambda,R_0)\cos\left(\frac{2\pi}{\Lambda} x\right)\right.\nonumber\\
&\left.+A_{\parallel}(\Lambda/2,R_0)\cos\left(\frac{4\pi}{\Lambda}x+\delta\right)\right],
\end{align}
where $P$ is the total optical power, $c$ represents the light speed in vacuum, $\delta$ is a relative phase shift between the two patterns, and the period of the lattice is given by $\Lambda=\lambda/2\sin\beta$ (see Fig.~\ref{fig_3beams}). The amplitude of the optical force for each of the overlapped light lattices is determined by the coefficients $A_{\perp}(\Lambda, R_0)$ and $A_{\parallel}(\Lambda/2, R_0)$, associated with polarization planes normal ($\perp$) and parallel ($\parallel$) to the incidence plane, defined by the sample and the direction of propagation of the beam. For the sake of clarity, their dependence on the period of the respective light pattern and on the radius of the particle is shown explicitly. A negative (positive) sign of these coefficients indicates that the particle finds stable equilibrium positions at the intensity maxima (minima) of the light lattice. 
 
There are different approaches for the calculation of the coefficients $A_{\perp}$ and $A_{\parallel}$; we use here a ray tracing model, described in the Appendix, which was experimentally validated in a previous work \cite{arzola2009force}. The optical potential is obtained by direct integration of Eq.~\eqref{eq_fuerza}, yielding 

\begin{equation}\label{eq_potencial}
V(x;\Lambda,R_0)=-V_0\left[\sin\left(\frac{2\pi}{\Lambda} x\right)+\frac{K}{2}\sin\left(\frac{4\pi}{\Lambda} x+\delta\right)\right],
\end{equation}
where $V_0=\lvert A_{\perp}(\Lambda,R_0)\rvert P \Lambda/2\pi c$ and 
\begin{equation}\label{eq_Kdef}
K(\Lambda/2,R_0)=\dfrac{A_{\parallel}(\Lambda/2,R_0)}{\lvert A_{\perp}(\Lambda,R_0)\rvert}. 
\end{equation}
This is an important parameter associated with the asymmetry of the potential and we will discuss it in detail in the next section.

We have a full expression for the total force given by the optical field and the corresponding potential given in Eq.~\eqref{eq_potencial}. This potential is the so called ratchet potential, used by many authors in the literature \cite{mateos_chaotic_2000,hanggi_artificial_2009}. This is a periodic and asymmetric potential that we obtained experimentally. 

\subsection{Dynamic rocking ratchet}

\begin{figure}
 \includegraphics[angle=0,width=3in]{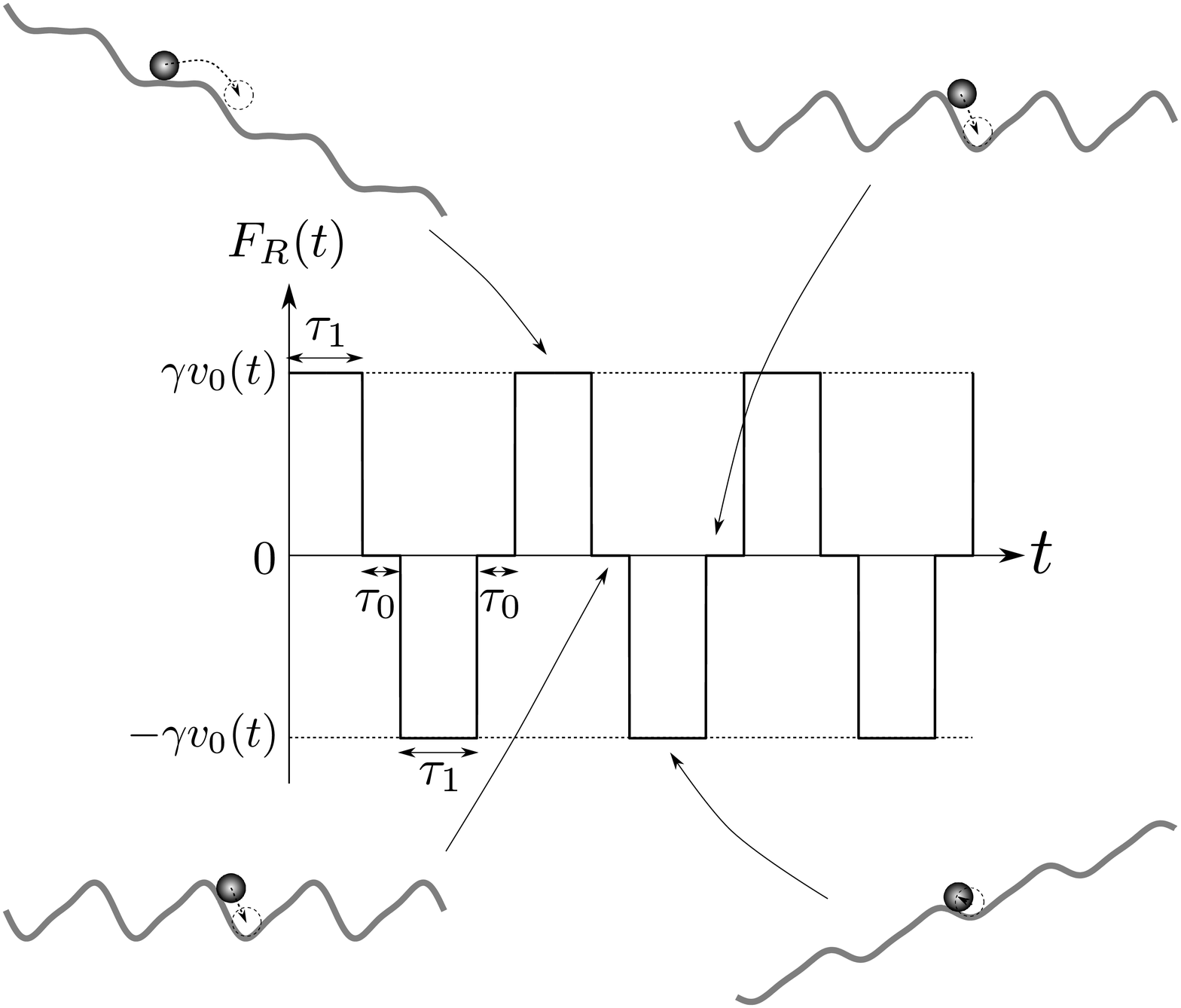}
\caption{Rocking mechanism of the ratchet potential. The figure shows the temporal dependence of the rocking force, and how when the time is running the asymmetric potential is tilted in one and in the opposite direction, given rise to a rectification process. It is important to stress that in our case we have the waiting-time $\tau_0$, which lets the particle go down into the minimum of the potential after each tilting-time $\tau_1$.}\label{fig_dynamics}
\end{figure}

In order to obtain a finite current or transport of particles we need to apply an external forcing to this ratchet potential  and, to obtain the non-trivial ratchet effect, this time-dependent external force should have zero average. A schematic of the rocking mechanism is depicted in Fig.~\ref{fig_dynamics}, where we show the external rocking force as a function of time, where we clearly see the periodicity and its unbiased character. The net effect of having a constant force is to tilt the ratchet potential where this tilt can have zero, positive or negative slope, as is illustrated in the figure. Due to the asymmetry of the ratchet potential, this tilting mechanism can induce a finite transport in one direction.          

In general, the Brownian motion of particles with radii larger than $3\mu m$ immerse in water at room temperature can be neglected \cite{arzola2009force}. The effect of thermal noise can be quantified by comparing both the thermal energy with the potential barriers, and the characteristic displacement due to diffusion with the characteristic length of the system, which in our case is the period of the optical lattice ($\Lambda$). In our system, the barrier of the potential energy is about $V_{max} \sim 1pN(\Lambda/2\pi)$. On the other hand the period of the optical lattice is of the order of $\Lambda\sim10\mu m$. Therefore, for room temperature, $kT/V_{max}\sim 10^{-3}$, and thus the thermal energy is negligible in comparison with the optical potential energy. On the other hand, the diffusion length of a particle with diffusion coefficient $D$ is $l^*=\sqrt{2Dt}$. Assuming that the characteristic time is of the order of one second and using the Stokes drag coefficient to estimate the diffusion $D=kT/6\pi\eta R_0$ (where $R_0$ is the radius of the particle taken as $3\mu m$ and $\eta$ is the viscosity of water), we have that $l^*/\Lambda\sim10^{-2}$. Thus, it is possible to neglect the effects of both the thermal diffusion and the thermal activation and, therefore, our system can be described by a deterministic and overdamped dynamics. The equation of motion is given by:

\begin{equation}\label{eq_modelo}
\gamma \dot{x}=-\frac{\partial}{\partial x} V(x)+F_R(t),  
\end{equation}
with $\gamma$ representing an effective drag coefficient \cite{arzola2009force, arzola2011experimental}, and $V(x)$ is given by equation \eqref{eq_potencial}. The time-periodic rocking force $F_R(t)$ is in this case a drag force owed to an oscillatory motion of the sample stage relative to the static periodic pattern of light, \textit{i.e.} $F_R(t)=\gamma v(t)$ \cite{arzola2011experimental}. The speed $v(t)$ is given by:

\begin{equation}\label{eq_modgamma}
v(t)=
\begin{cases}
v_0 & \text{if}\ \ 0\leq t< \tau_1\\
0 & \text{if}\ \ \tau_1\leq t< \tau_1+\tau_0\\
-v_0 & \text{if} \ \ \tau_1+\tau_0\leq t< T-\tau_0\\
0 & \text{if} \ \ T-\tau_0\leq t< T          
\end{cases}
\end{equation}
with $v_0$ a constant speed. The time $\tau_0$ is the waiting-time and $\tau_1$ is the tilting-time. It is important to stress that the time-average of $F_R(t)$ over a period, $T=2(\tau_0+\tau_1)$, is zero in order to have an unbiased forcing, and thus, the nontrivial ratchet transport (Fig.~\ref{fig_dynamics}). As we shall see, the role of the waiting time $\tau_0$ is very important in the dynamics of the system. 

In summary, in this section we have described the experimental setup of an optical rocking ratchet, and in the rest of this paper we will analyze the corresponding theoretical model, the role of the asymmetry and a comparison between theory and experiment.

\section{Control of the asymmetry of the ratchet potential: a useful tool for particle fractionation}
\begin{figure*}
\includegraphics[angle=0,width=6in]{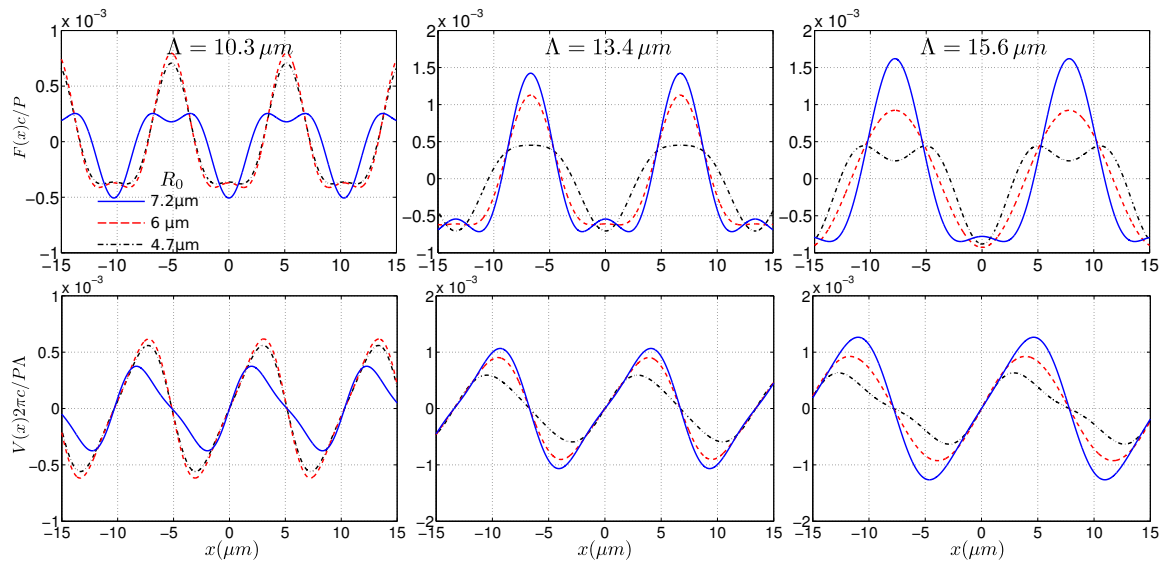}
\caption{(Color online) Calculated optical force (top row) and optical potential (bottom row) as functions of $x$ for three different particles with radii of $7.2\,\mu m$ (solid curves), $6\,\mu m$ (dashed) and $4.7\,\mu m$ (dash-dot), interacting with asymmetric light patterns  of different periods: $10.3\,\mu m$ (left), $13.4\,\mu m$ (center) and $15.6\,\mu m$ (right). In all cases $\delta=0$. Both the force and the potential have been normalized to be expressed as dimensionless quantities.}\label{fig_forces}
\end{figure*}

We will analyze the role of the symmetry of the ratchet potential, Eq.~\eqref{eq_potencial}, in the generation of transport. Additionally to the parameter $K$ in this potential and in the optical force, we have the control parameter $\delta$, which represents a phase shift, as discussed in Ref. \cite{arzola2011experimental}. Both parameters,  $K$ and $\delta$, determine the shape of the curves for the total force and potential as a function of $x$, as is illustrated in Fig.~\ref{fig_forces} for some representative cases. For instance, the typical ratchet potential is obtained by setting $K=0.5$ and $\delta=0$ in Eq.~\eqref{eq_potencial} \cite{mateos_chaotic_2000,hanggi_artificial_2009}, but its asymmetry is inverted if $K=-0.5$ or $\delta=\pi$. In general, negative and positive values of $K$ for $\delta=0$ give rise to potentials with opposite asymmetries, and the same occurs for a fixed $K$ when we replace $\delta=0$ by $\delta=\pi$. On the other hand, for $\delta=\pm\pi/2$ the potential is symmetric regardless of the value of $K$. These two parameters have  experimentally different character: while $K$ is fixed, once the optical lattice and the dielectric spheres are chosen, the phase shift $\delta$ can be controlled and changed at will in real time.

With the aim of comparing the asymmetry of the potentials for different particles, we found it very useful to define an \textit{asymmetry parameter} as 

\begin{equation}\label{eq_Ia}
\alpha=2\left( \frac{|F_{max}|-|F_{min}|}{max(|F_{max}|,|F_{min}|)}\right). 
\end{equation}

\begin{figure}
\includegraphics[angle=0,width=3.5in]{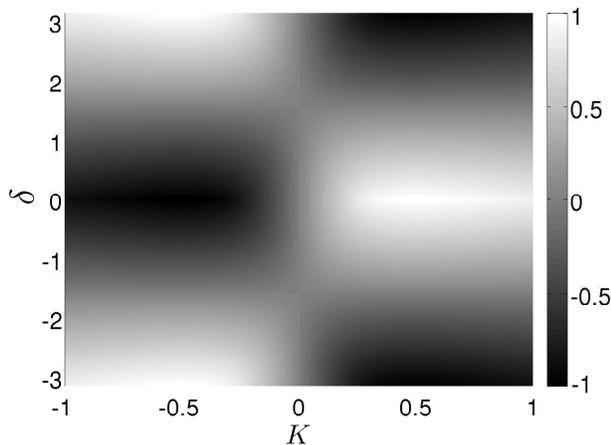}
\caption{Asymmetry parameter, $\alpha$, as a function of $K$ and $\delta$ (in radians).}\label{fig_asim}
\end{figure}

This dimensionless parameter, which varies between $0$ and $1$, measures the difference between the maximum ($F_{max}$) and the minimum ($F_{min}$) values of the force as a function of $x$, as seen in Fig.~\ref{fig_forces}.  These extreme values correspond to the values of the slope in the ratchet potential. Notice that $|F_{min}|$ can be larger than $|F_{max}|$, and therefore the parameter $\alpha$ can be either positive or negative (see Fig.~\ref{fig_forces}). The case when $\alpha=0$ corresponds to a symmetric periodic potential. This parameter is depicted in Fig.~\ref{fig_asim} as a function of $K$ and $\delta$. Notice as well that when $K=\pm0.5$, and $\delta=0$, $\alpha=\pm1$ which corresponds to the extreme case when the asymmetry of the ratchet potential is maximal; this case can be obtained as well for $K\mp0.5$ and $\delta=\pi$. On the other hand, the case $\alpha=0$ is obtained for $\delta=\pm\pi/2$ for all values of $K$, and for $K=0$ for all values of $\delta$, corresponding to a symmetric force and potential. For the case of $\delta=0$ is easy to show that the \textit{asymmetry parameter} is given by $\alpha=(2|K|-1/4)/K(1+|K|)$, and its maximum corresponds to $K=\pm1/2$.

\begin{figure}
\includegraphics[angle=0,width=3in]{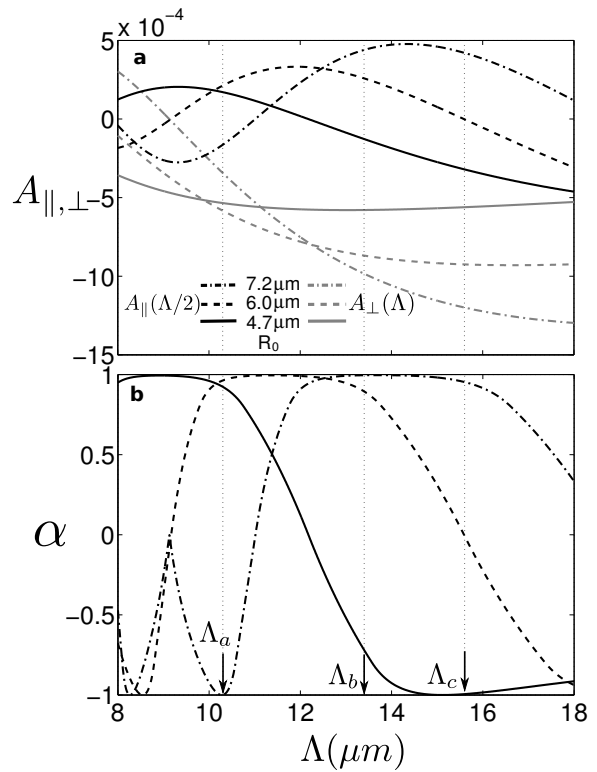}
\caption{(a) Coefficients of the calculated optical force for each of the two superimposed light patterns of fringes with periods $\Lambda$ (gray curves) and $\Lambda/2$ (black curves) as a function of $\Lambda$, for spheres of radii $4.7$ (solid curves), $6$ (dashed) and $7.2\,\mu m$ (dash-dot). (b) Asymmetry parameter $\alpha$ against the period $\Lambda$ for the same diameters.}\label{fig_coefficients}
\end{figure}

In Fig.~\ref{fig_coefficients}a, we show the coefficients $A_\parallel$ and $A_\perp$ of Eq.~\eqref{eq_fuerza} as a function of $\Lambda$ for the three different sizes we used in our experiments. In Fig.~\ref{fig_coefficients}b we depict the parameter $\alpha$ for the optical force as a function of $\Lambda$; in this plot we show the three periods illustrated in Fig.~\ref{fig_forces} denoted as $\Lambda_a$, $\Lambda_b$ and $\Lambda_c$. The sign of $A_\parallel$ can be positive or negative and, according to Eq.~\eqref{eq_Kdef}, the sign of the parameter $K$ in the ratchet potential can be positive or negative as well, as seen in Fig.~\ref{fig_coefficients}b. This means that we can control the asymmetry of the ratchet potential by varying $K$, as shown in Fig.~\ref{fig_forces}. By analyzing the sign of $\alpha$ in Fig.~\ref{fig_coefficients}b, we can infer the asymmetry of the ratchet potential and associate it with the cases illustrated in Fig.~\ref{fig_forces}. In the particular case of $K=0$, the potential becomes symmetric so we do not expect to obtain a transport of particles. 

\begin{figure}
\includegraphics[angle=0,width=3.5in]{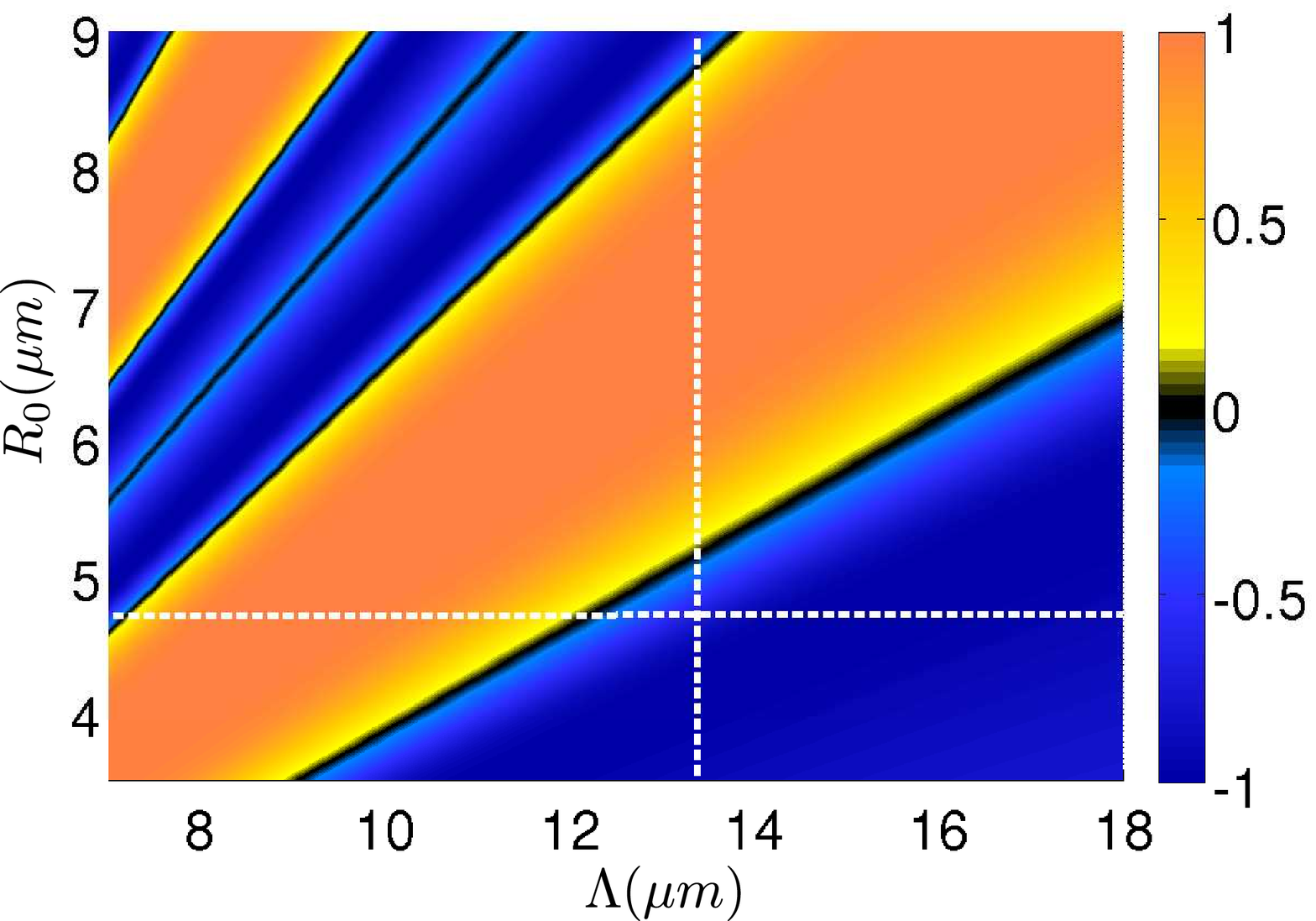}
\caption{(Color online) Asymmetry parameter $\alpha$ as a function of the period $\Lambda$ and the radius of the particle $R_0$.}\label{fig_alpha}
\end{figure}

Fig.~\ref{fig_alpha}, in addition, illustrates the parameter $\alpha$ for the optical force, as a function of the lattice period $\Lambda$ and of the radius of the particle $R_0$. It is worth mentioning that, even for the smallest particle radius analyzed here ($R_0\approx3.5\,\mu m$), the ray tracing model used to calculate the optical force is still reasonable (see ref. \cite{arzola2009force}). The value of $R_0=4.7\mu m$ is illustrated with a dashed horizontal line in Fig.~\ref{fig_alpha} as an example since this is one of the sizes of microspheres we used in our experiment. It is seen that $\alpha>0$ in the range of periods $7.2\,\mu m \lesssim \Lambda \lesssim 12.2\,\mu m$, while $\alpha<0$ for $\Lambda\gtrsim 12.2\,\mu m$. Therefore, we can vary the asymmetry of the optical ratchet potential for a fixed size of a microsphere by changing the period of the optical lattice. Now consider a fixed period $\Lambda=13.4\,\mu m$ (vertical dashed line), corresponding as well to one of the experimental values, $\alpha>0$ for spheres of radii within the range $5.15\,\mu m\lesssim R_0\lesssim 8.75\,\mu m$, while $\alpha<0$ for $ R_0\lesssim5.15\,\mu m$ or $R_0\gtrsim8.75\,\mu m$. Therefore, from this diagram it is easy to identify those particles whose radii are in the ranges where $\alpha>0$, which will experience a potential with opposite asymmetry with respect to those for which $\alpha<0$. This implies that it is possible to observe simultaneous opposite motion of particles of different sizes within the same light pattern in our ratchet system, as it was demonstrated in \cite{arzola2011experimental}, or to invert the motion of a given particle by changing the period of the light lattice.

Furthermore, the diagram of Fig.~\ref{fig_alpha} can be used as a very good starting-point criterion for particle sorting or fractionation, since it can be used to determine a period for which the asymmetry of the optical potential experienced by two species of particles within the same light pattern is opposite. In that case, it should be always possible to find a set of dynamical parameters, regarding the magnitude and time-period of the external driving force, that allow the separation. For instance, one can wonder which period of the optical potential would be appropriated in order to separate particles with radii of $6\,\mu m$ and $6.6\,\mu m$, which differ only by $10\%$. As it can be seen from Fig.~\ref{fig_alpha}, for a period of $\Lambda=9.7\,\mu m$, $\alpha=0.73$ for $R_0=6\,\mu m$ and $\alpha=-0.77$ for $R_0=6.6\,\mu m$, which is expected to give rise to a successful separation.

Nevertheless, as we shall see in the next section, in order to determine the behavior of a given particle in a rocking ratchet, it is necessary to perform a full dynamic analysis of the system.

\section{Dynamical model of a deterministic optical rocking ratchet}
In order to perform a more general analysis, it is convenient to rewrite Eq.~\eqref{eq_modelo} in terms of dimensionless  variables, namely,  

\begin{equation}\label{eq_modelodimen}
\frac{d}{d\tilde{t}}\tilde x(\tilde{t})=-\frac{\partial}{\partial \tilde x} \tilde V(\tilde x)+\tilde f(\tilde t),  
\end{equation}
where $\tilde{x}=\frac{x}{\Lambda}$, $\tilde{t}=\frac{F_0}{\gamma\Lambda}t$ and $\tilde{f}=\frac{F_R(t)}{F_0}$, with $F_0=2\pi V_0/\Lambda$. The dimensionless optical potential is now expressed as
\begin{equation}
\tilde{V}(\tilde{x})=-\frac{1}{2 \pi}\left[\sin(2\pi\tilde{x})+\frac{K}{2}\sin(4\pi\tilde{x}+\delta)\right].
\end{equation}
Notice that the asymmetry parameter $\alpha$ will remain unaltered.     

The main quantity that characterizes the dynamics of the ratchet effect is the current, which is the asymptotic mean velocity of the particle in the periodic ratchet. We define the dimensionless current $J$ as the number of spatial periods that the particle traverses during a full period of time $T$ of the external forcing.   

\begin{figure}
\includegraphics[angle=0,width=3.2in]{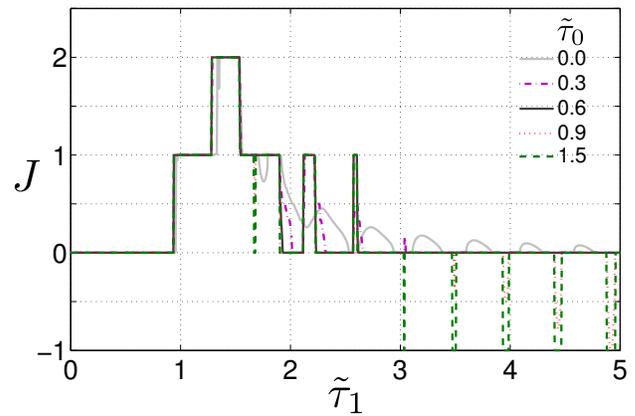}
\caption{(Color online) Normalized current $J$ as a function of the dimensionless force $\tilde{f}$, calculated from the equation \eqref{eq_modelodimen} for four different values of the waiting-time $\tilde{\tau}_0$: $0$ (solid gray curve); $0.3$ (dash-dot); $0.6$ (solid black); $0.9$ (dotted) and $1.5$ (dashed). In all the cases the tilting-time was $\tilde{\tau}_1=2$.}\label{figdiagrama3}
\end{figure}

Now we solved the differential equation \eqref{eq_modelodimen} numerically and obtain the current $J$ after a large number of periods $T$. In Fig.~\ref{figdiagrama3} we show $J$ as a function of $\tilde{f}$ for four different values of the waiting-time $\tilde{\tau}_0$, for a fixed value of the parameter $K=0.5$. Here the role of the waiting-time $\tilde{\tau_0}$ becomes very relevant. 

When $\tilde{\tau_0}\rightarrow 0$, the system may exhibit a chaotic behavior. This is illustrated in Fig.\ref{figdiagrama3}, where we show four characteristic curves for the current $J$ as a function of $\tilde{f}$ for different values of the waiting-time $\tilde{\tau_0}$, for $K=0.5$. As described by Zarlenga and coworkers~\cite{zarlenga_complex_2009}, the plateau regions correspond to periodic trajectories of the particle, while the humps, appearing for low values of $\tilde{\tau}_0$, are characterized by the chaotic motion of the particle. This regime was not accessible with our experimental setup due to technical limitations of the motor driving the sample stage. As the value of $\tilde{\tau}_0$ grows, the humps turn narrower and weaker and new plateaus, \textit{i.e.} new periodic trajectories, arise. For the highest values of $\tilde{\tau}_0$ the current takes only integer values, as it is our case. This perfect periodic behavior is owing to the fact that $\tilde{\tau}_0$ is long enough to allow the particle to reach a stable equilibrium position in the spatial potential after each activation semicycle $\tilde{\tau}_1$, then the particle starts every new cycle with the same initial conditions regarding its relative position in a potential well. In that case, the current can be expressed as $J_N=n-m$, where $n$ ($m$) is the number of periods that the particle is able to move along the direction of lowest (highest) slope in the potential in each semicycle. The values of $n$ and $m$ depend on $\tilde{\tau}_1$ and $\tilde{f}$ for a given potential. An important behavior shown in Fig.~\ref{figdiagrama3}, first described by Reimann \textit{et al.} \cite{Reimann_giant_1998} and recently experimentally showed by the authors \cite{arzola2011experimental}, is the current reversals that emerges when $\tilde{\tau}_0$ is different from zero.

\begin{figure}
\includegraphics[angle=0,width=3.4in]{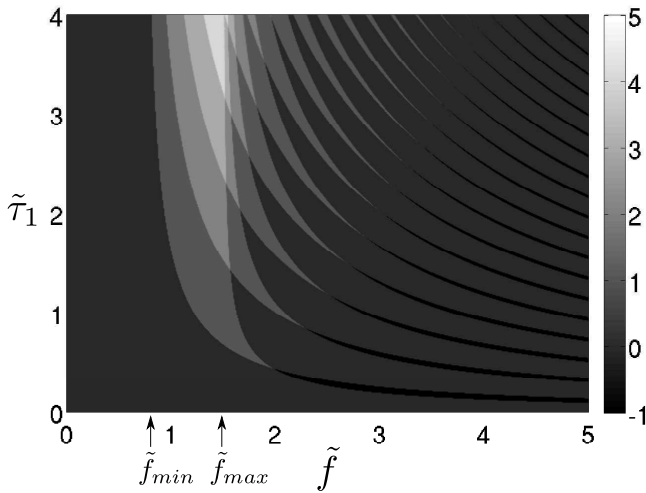}
\caption{\label{figdiagrama1}Normalized current (gray scale) as a function of the dimensionless force $\tilde{f}$ and the activation tilting-time $\tilde{\tau}_1$, according to the equation \eqref{eq_modelodimen}. The waiting-time is $\tilde{\tau}_0=1.5$.}
\end{figure}

Figure \ref{figdiagrama1} depicts the complete diagram of parameters of the current (grayscale level) against $\tilde{f}$ and $\tilde{\tau}_1$, for $\tilde{\tau}_0=2$. For $\tilde{f}\leq\tilde{f}_{min}$ the particle is not able to surmount any energy barrier, so the current is obviously zero. When $\tilde{f}$ is larger than $\tilde{f}_{min}$ a characteristic stair-like behavior appears \cite{hanggi_artificial_2009,Reimann_giant_1998,alatriste_phase_2006}. For $\tilde{f}_{min}<\tilde{f}\leq\tilde{f}_{max}$, the particle can climb out from the potential well only along the lower slope (positive direction in this example), so for a fixed $\tilde{f}$ the current increases by one when the activation time $\tilde{\tau}_1$ increases by an amount equal to the time needed to go from one maxima to the next one. For $\tilde{f}>\tilde{f}_{max}$ the current is the result of the combination of forward and backward displacements, so the rectification mechanism turns less efficient as we increase the force. In this regime the value of the current is determined by the time intervals that the particle takes to go from one minima in the energy potential to a maxima in forward and in backward directions, in the corresponding activation semicycles. The current reversals, which always have a negative value  $J_N=-1$, cover very narrow regions where the particle can jump $n$ periods in forward direction and $n+1$ in the backward direction. These regions have the peculiarity that the particle takes less time to climb the steeper slope of the well (maximum opposite force) than the lower slope (minimum opposite force). This effect is allowed only when the waiting time $\tilde{\tau}_0$ is different from zero. The points where the current changes its direction in the diagram correspond to the conditions where the particle takes the same time climbing out from the potential well in both directions. 

\section{Experimental demonstration of current reversals}

\begin{figure*}[!ht]
\includegraphics[scale=0.42]{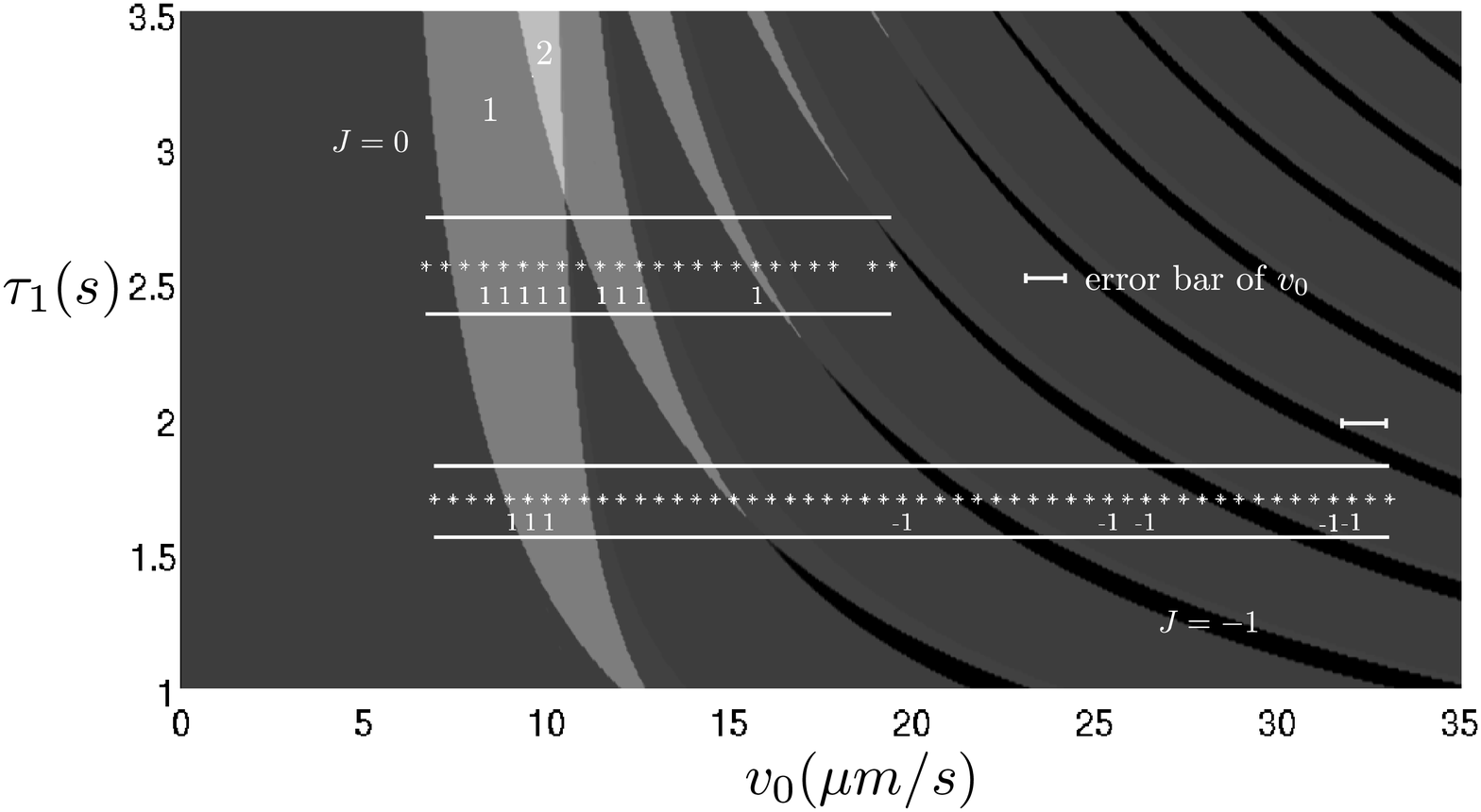}
\caption{Comparison between experimental and theoretical values of the current $J$ as a function of $v_0$ and $\tau_1$. Marked with asterisk, two series of experimental data with different tilting-time are shown. Top: $\tau_1=(2.56\pm0.18)\,s$; bottom: $\tau_1=(1.70\pm0.13)\,s$. The white lines above and below the experimental data indicate the uncertainty of $\tau_1$. In all cases $\tau_0=(2.00\pm0.05)\,s$. The value of the current for the experimental points is zero unless otherwise indicated. The radius of the particle is $R_0=(7.25\pm0.30)\,\mu m$ and the asymmetric light pattern is characterized by the parameters: $\Lambda=(13\pm0.1)\,\mu m$, $P=(1.63\pm0.05)\,W$, $W_x=(745\pm5)\,\mu m$, $W_y=(19\pm2)\,\mu m$ and $\delta=0$. The effective friction coefficient considered for the calculation was $\gamma=0.359 pN.s/\mu m$.}\label{figexpteo}
\end{figure*}

\begin{figure}[!ht]
\includegraphics[scale=0.4]{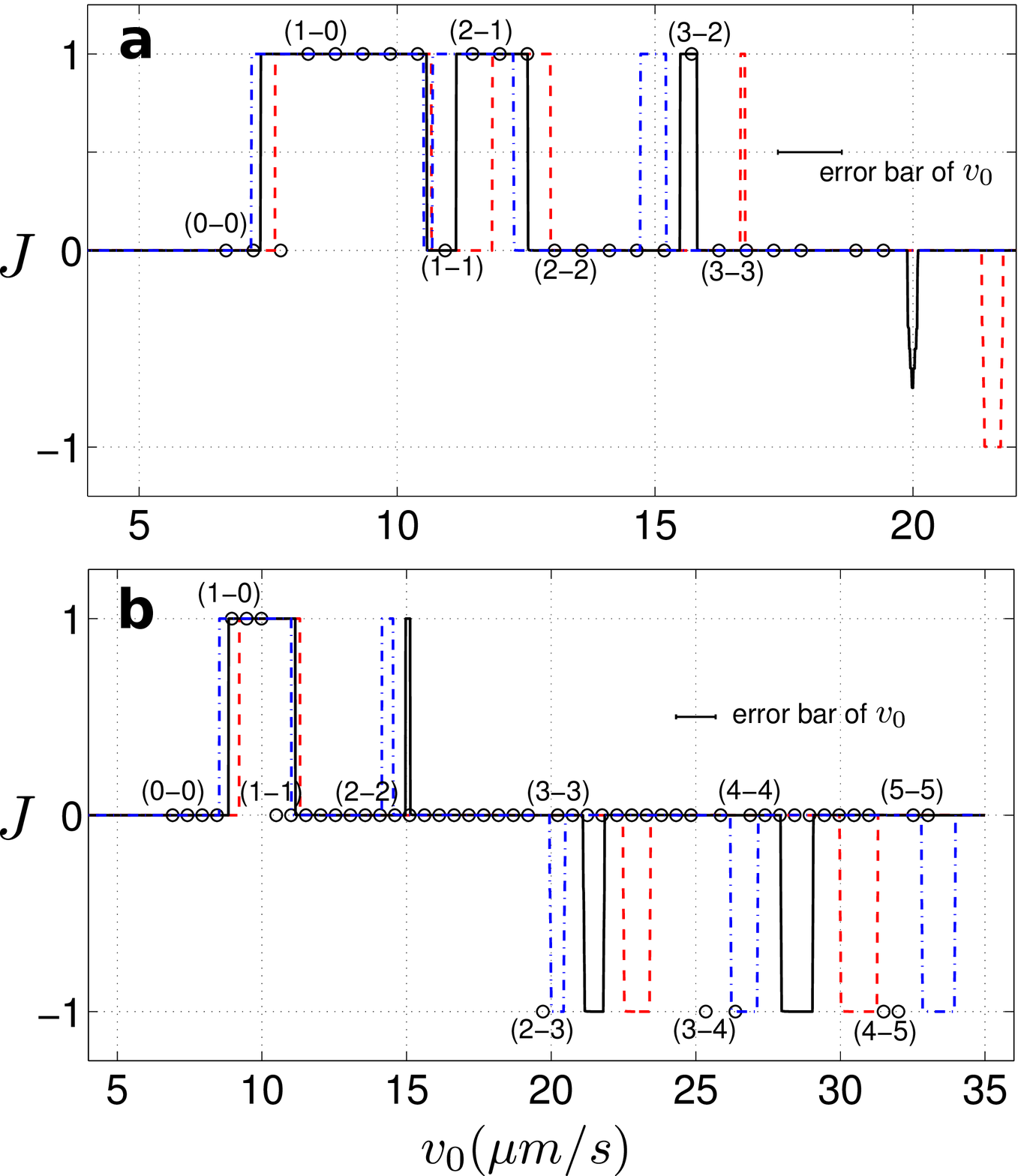}
\caption{\label{fig_expteo2} Experimental data sets shown in Fig.~\ref{figexpteo} for $J$ as a function of $v_0$ (circle markers), corresponding to: a) $\tau_1=(2.56\pm0.18)s$ and b) $\tau_1=(1.70\pm0.13)s$. The values $n$ and $m$ that leads to the current ($J=n-m$) are written next to the experimental points. The curves represent the theoretical calculated values for the experimental data of $\tau_1$ (solid curves), the maximum value of $\tau_1$ within the uncertainty interval (dash-dot) and its minimum value (dashed).}
\end{figure}

Having described the theoretical background of the experiment that leads to an optical ratchet system, in this last section we will make a quantitative comparison of theory and experiment. In Fig.~\ref{figexpteo} we can see two series of experimental data superimposed on the theoretical diagram of parameters. The radius of the particle is $R_0=(7.25\pm0.30)\,\mu m$ and the period of the asymmetric pattern is $\Lambda=(13.0\pm0.1)\,\mu m$. The value of the experimental current is indicated below every point, and the white lines above and below the experimental points indicate the experimental uncertainty along the vertical scale (tilting-time $\tau_1$), while the uncertainty of $v_0$ is depicted in the left side of the figure as an error bar. Notice that in this case we are plotting the data with physical units, so we are not using the dimensionless variables anymore. These two experimental series are shown in Fig.~\ref{fig_expteo2} as a function of the velocity $v_0$, and next to each experimental data, the values $n$ and $m$ that leads to the current $J=(n-m)$ are written. It is important to stress that there are no fitted values in the comparison. The parameters required for the theoretical comparison, such as the radius $R_0$, the laser power into the sample $P$, the beam waist $w_x$ and $w_y$, and the period $\Lambda$, were independently determined. The numerical values are indicated in the caption of the experimental plot. The refractive indices assumed for the particles and the water were, respectively, $n_p=1.56$ (Duke Scientific, Borosilicate microspheres) and $n_m=1.333$. The effective friction coefficient $\gamma$ was determined indirectly with a method described in detail in ref. \cite{arzola2009force}.

\begin{figure}
\includegraphics[scale=0.4]{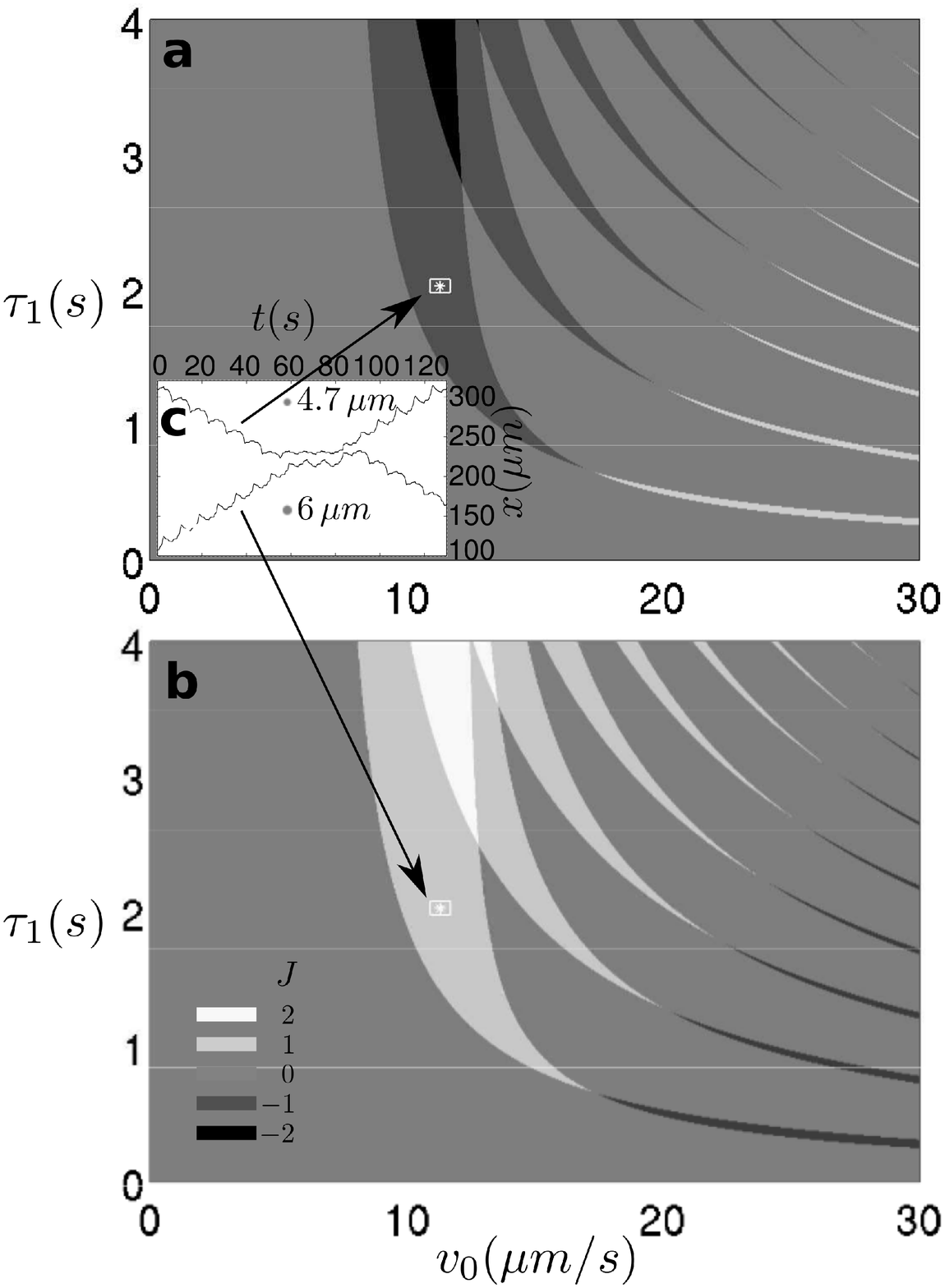}
\caption{Example of the size-dependent direction of the current. Two experimental currents corresponding to two different size particles are contrasted with its corresponding theoretical diagram of parameters; a) $R_0=(4.70\pm0.15)\,\mu m$ and b) $R_0=(6.00\pm0.15)\, \mu m$.  The experimental parameters are: $P=(1.67\pm0.05)W$, $v_0=(11.3\pm0.4)\mu m/s$, $\tau_0=(2.00\pm0.05)s$, $\tau_1=(2.03\pm0.05)s$, $\Lambda=(13.4\pm0.1)\mu m$. Fig. c) shows the trajectories of the these two particles when these are subject to the mentioned conditions. Notice that the colormap for J is the same for both plots.}\label{fig:size} 
\end{figure}

As we stated before, the magnitude and shape of the energy potential depends appreciably on the size of the particle, so it follows that the dynamics of two different particles within the same light pattern will present very different behaviors. To illustrate this effect experimentally, we placed two particles with different radii ($R_0=(4.70\pm0.15)\,\mu m$ and $R_0=(6.00\pm0.15)\, \mu m$) at a time in our optical ratchet system. We chose the period of the optical lattice in such a way to obtain a high asymmetry (see Fig.~\ref{fig_coefficients}), and for some combination of parameters $\tau_1$ and $v_0$, for fixed power P, we observed that the average currents were in opposite directions. In Figs.~\ref{fig:size} (a)- (b), the average current for these two particles, as a function of $\tau_1$ and $v_0$, is contrasted. These two diagrams share the same general properties discussed before (Fig.~\ref{figdiagrama1}), but they are quantitatively very different; while the current of the particle with radius $R_0=6.00\,\mu m$ is dominated by positive values, the particle with radius $R_0=4.70\,\mu m$ follows currents with opposite sign predominantly. An experimental point of the current for these conditions, taken from \cite{arzola2011experimental}, is shown in the plots.

\section{Conclusions}

Based on the system introduced in a previous work \cite{arzola2011experimental}, we presented here an extended analysis and new results on the dynamics of a deterministic optical rocking ratchet. Firstly, with the aim of analyzing the particle-dependence of the optical potential, we computed the conservative contribution of the optical force exerted by a superposition of two periodic light patterns on a spherical dielectric particle using a ray optics model, suitable to describe the experimental conditions. We showed that, although the resulting light pattern is asymmetric in the absence of the particle, the light-particle interaction gives rise to a modified optical potential, whose symmetry depends on the ratio between the particle size and the period of the lattice. This modified potential can have either the same asymmetry of the light pattern, the opposite one, or it can even be symmetric. Therefore, in order to characterize the optical potential in a simple way, we defined an ``asymmetry parameter''  $\alpha$, such that $\alpha=\pm1$ corresponds to maximum and opposite asymmetries, and $\alpha=0$ stands for a symmetric potential. Moreover, this parameter was used to predict, in a first approach, the direction of the current in the ratchet system as a function of particle size and period, which would be very useful to design a sensitive fractionation device. In that respect, it is worth to stress that our interest here was only to present a probe of principle. A useful sorting device would require, of course, an extended light pattern in the sample plane, in contrast to the quasi-1D pattern that we generated in our experiment. However, this does not represent a big experimental challenge.   

On the other hand, we presented a thorough analysis of the complex dynamics of the ratchet system. For a given ratchet potential, we analyzed the current of the system in terms of the three parameters defining the rocking force: its magnitude, the tilting-time and the waiting-time. Here it was shown that the waiting-time is a necessary condition for the arising of current reversals, observed experimentally for the first time in Ref.~\cite{arzola2011experimental}. In fact, this phenomenon had been predicted theoretically for a similar system since 1998 \cite{Reimann_giant_1998, salgado-garcia_deterministic_2006}. 

In contrast, as the waiting-time decreases, the current reversals start to disappear and, in the limit when it tends to zero, a chaotic behavior might be expected \cite{zarlenga_complex_2009}. Although this regime was not accessible in our actual experiment due to technical limitations of the step motor driving the sample stage, our setup allows future experiments to explore this problem by replacing the motor with a more versatile and accurate device, like a piezoelectric actuator.

In the last section, we applied the theoretical model to describe some specific experimental situations. A quantitative comparison of the respective results for the current was established for a given particle size and light lattice period, by varying the magnitude and tilting-time of the rocking force. We found good agreement of the general behavior within the experimental uncertainty, and importantly, there were no fitted values in this comparison. Furthermore, an explicit example of the comparative dynamics for two different particles within the same light pattern allowed to appreciate the role of the opposite asymmetry of the optical potentials in the dynamics.

By means of our experimental system we were able to demonstrate the interesting phenomenon of current reversals. However, there are several aspects of the ratchets systems that remain open problems, such as the study of the chaotic regime \cite{zarlenga_complex_2009}, the role of interactions between particles \cite{malgaretti_running_2012, liebchen_interaction-induced_2012}, the Brownian rocking ratchets, among many others. Optical micromanipulation techniques have proved to be very versatile tools for investigating novel aspects of non-linear dynamical systems and validate theoretical models in a crystal clear manner.  
 
\begin{acknowledgments}
This project was partially supported by DGAPA-UNAM (grant IN100110), CONACYT Mexico (grant 132527), CONACYT-ASCR Mexico-Czech Republic bilateral cooperation project (grant 171478). A. V. A also acknowledges support from MEYS CR (project No. LH12018) and EC and MEYS CR (project No. CZ.1.07/2.4.00/31.0016).
\end{acknowledgments}

\section{appendix}
Here we will present a brief summary of the ray tracing model used for the calculation of the optical forces. Although the applicability of this model is limited to the case of large particles compared with the wavelength, it is important to point out that the general results presented in this paper would be still valid if the calculation of the optical forces are performed with a different model. In our experimental conditions, however, the ray tracing model has probed to be an appropriated approximation \cite{arzola2009force}. 

The $x$ component of the optical force exerted on a spherical particle, located at the position $(x_0,y_0)$ in a transverse plane ($z = \text{constant}$), by a collimated light field with transverse intensity distribution $I(x,y)$ and propagating vertically upwards can be written as,
\begin{multline}  \label{forcemod}
F_x (x_0,y_0)=-\frac{(R_0)^2 n_m}{2c} \int_0^{\pi/2} \int_0^{2\pi}
\vphantom{\frac{1}{2}}M(\theta)\sin(2\theta)\\
\cdotp I(x,y)\cos\phi\,d\phi%
\,d\theta,
\end{multline}
with

\begin{align}
M(\theta)=\mathfrak{R}\sin(2\theta)-\mathfrak{T}^2\frac{\sin(2\theta-2\theta_t)+\mathfrak{R}\sin(2
\theta)}{1+\mathfrak{R}^2+2\mathfrak{R}\cos(2\theta_t)}\nonumber,
\end{align}
where $c_m=c/n_m$ is the light speed in the host medium, and the integration is performed over the illuminated hemisphere of the particle, with $\phi$ and $\theta$ denoting the azimuthal angle and the angle complementary to the polar angle, respectively. The incidence angle $\theta_{i}$ coincides with $\theta$ at each point on the sphere's surface, and $\theta_t$ is the transmitted angle, and is given by Snell's law. The reflectance $\mathfrak{R}$ for a linearly polarized beam, whose polarization plane forms an angle $\mu$ with respecto to the $x$ direction is 
\begin{equation}\label{R}
\mathfrak{R}_\mu=\cos^{2}\left( \mu-\theta_{i}\right) \mathfrak{R}_{1} + \sin^{2}\left( \mu-\theta_{i}\right) \mathfrak{R}_{2}, 
\end{equation}
with 
\begin{equation}\label{Rpl}
\mathfrak{R}_{1}=\frac{\tan^{2}\left( \theta_{i}-\theta_{t}\right) }{\tan^{2}\left( \theta_{i}+\theta_{t}\right) }
\end{equation}
and 
\begin{equation}\label{Rpl}
\mathfrak{R}_{2}=\frac{\sin^{2}\left( \theta_{i}-\theta_{t}\right) }{\sin^{2}\left( \theta_{i}+\theta_{t}\right) },
\end{equation}
whereas the transmittance is simply given by $\mathfrak{T}_\mu=1-\mathfrak{R}_\mu$, neglecting absorption. In our case $\mu$ takes the values $0$ and $\pi/2$ for the incidence $\parallel$ and $\perp$, respectively. The coordinates of each point at the particle's surface, ($R_0 $, $\phi$, $\theta$) (with $R_0$ the radius of the microsphere), and the position of the center of the particle with respect to the beam axis, ($x_0$ $y_0$), are related by means of $x=x_0+R_0\cos\phi\sin\theta$ and $y=y_0+R_0\sin\phi\sin\theta$, (we set $y_0=0$). The assumption that the $x$ component of the optical force,
described by Eq.~\eqref{forcemod}, is conservative, is based on the fact that there is no propagation of the beam along this direction \cite{ashkin1992forces}.

The intensity distribution for a single interference pattern of period $\Lambda$ is 
\begin{equation}  \label{optmod}
I(x, y)=\frac{2 P}{\pi w_x w_y}e^{-2\left( \frac{x^2}{w_x^2}+\frac{
y^2}{w_y^2}\right)}\left[ \cos\left(\frac{2\pi}{\Lambda} x\right)+1\right],
\end{equation}
where $P$ denotes the incident optical power at the sample plane. We have a Gaussian envelope whose widths along the $x$ and $y$ directions are, respectively, $w_x = (745\pm5)\,\mu m$ and $w_y = (19\pm2) \,\mu m$. We are in a regime where $w_x \gg \Lambda$, $w_x \gg w_y$ and $w_x \gg 2R_0$. In addition, the dynamics of the particles is observed within the central region of the pattern (about $250\,\mu m$ long). Therefore, we can disregard the effect of the Gaussian envelope along the $x$ direction and consider that we have a 1D optical lattice of period $\Lambda$. 
On the other hand, the total intensity when we have a superposition of two interference patterns with orthogonal polarizations and a relative phase between them $\delta$, one of them having twice the period of the other in order to make an asymmetric optical potential, is \cite{arzola2011experimental}

\begin{align}\label{eq_Iasim}
I_T(x, y)=&\frac{2P}{\pi w_x w_y}e^{-2\left(\frac{x^2}{w_x^2}
+\frac{
y^2}{w_y^2}\right)}\left[\sin^2(\varphi)\cos\left(\frac{2\pi}{\Lambda} x\right)
\right.\nonumber\\ 
 &\left.+\cos^2(\varphi)\cos\left(\frac{4\pi}{\Lambda}x+\delta-\pi/2\right)+1\right].
\end{align}
In this case the pattern is formed by the interference by pairs of three beams, one of them polarized along the $x$ direction, another one polarized along the $y$ direction, and a third one with its polarization plane forming an angle $\varphi$ with respect to the $x$ direction, as illustrated in Fig.~\ref{fig_3beams}. The coefficients $\sin^2\varphi$ and $\cos^2\varphi$ are associated with the pola rization angle of the third beam, which was set as $\varphi=\pi/4$ in the set of experiments reported here.

Using the intensity given by the equation \eqref{eq_Iasim}, and using the approximation discussed above ($w_x>>R_0$) in the expression for the gradient force \eqref{forcemod}, one arrives at the equation \eqref{eq_fuerza},where the coefficients are given by:

\begin{align}\label{eq_amplitudes}
A_{\{\parallel,\perp\}}=\frac{R^2_0 n_m}{\pi w_x w_y} \int_0^{\pi/2} \int_0^{2\pi}
S_{\{\parallel,\perp\}}e^{-2 (R_0\sin\phi\sin\theta)^2/w_y^2}\nonumber\\
\cdotp \sin\left(\frac{2\pi}{\Lambda} R_0\cos\phi\sin\theta\right)\sin(2\theta)\cos{\phi}\,d\phi\,d\theta,
\end{align}
with
\begin{equation}
 S_{\{\parallel,\perp\}}=\mathfrak{R}_{\mu}\sin(2\theta)-\mathfrak{T}_{\mu}^2\frac{\sin(2\theta-2\theta_t)+\mathfrak{R}_{\mu}\sin(2\theta)}{1+\mathfrak{R}_{\mu}^2+2\mathfrak{R}_{\mu}\cos(2\theta_t)}.
\end{equation}

%
\end{document}